\documentclass[10pt,twocolumn,english]{revtex4}
\usepackage[T1]{fontenc}
\usepackage[latin9]{inputenc}
\usepackage[a4paper]{geometry}
\usepackage{graphicx}
\usepackage{amssymb}
\usepackage{esint}

\makeatletter

\providecommand{\tabularnewline}{\\}

\@ifundefined{textcolor}{}
{%
 \definecolor{BLACK}{gray}{0}
 \definecolor{WHITE}{gray}{1}
 \definecolor{RED}{rgb}{1,0,0}
 \definecolor{GREEN}{rgb}{0,1,0}
 \definecolor{BLUE}{rgb}{0,0,1}
 \definecolor{CYAN}{cmyk}{1,0,0,0}
 \definecolor{MAGENTA}{cmyk}{0,1,0,0}
 \definecolor{YELLOW}{cmyk}{0,0,1,0}
 }

\makeatother

\usepackage{babel}

\begin{document}

\title{Scaling Solutions in Continuous Dimension}

\author{Alessandro Codello%
\thanks{codello@sissa.it%
}\\
}

\address{SISSA - Via Bonomea 265 - 34136, Trieste - Italy}
\begin{abstract}
We study scaling solutions of the RG flow equation for $\mathbb{Z}_{2}$-effective
potentials in continuous dimension $d\geq2$. As the dimension is lowered
from $d=4$ we first observe the appearance of the Ising scaling
solution and successively the apparence of multi-critical scaling solutions
of arbitrary order. Approaching $d=2$ these multi-critical scaling
solutions converge to the unitary minimal models found in CFT. 
\end{abstract}
\maketitle

\paragraph*{Introduction.}

One of the most valuable properties of the functional RG formalism
based on the effective average action \cite{Berges_Tetradis_Wetterich_2002} is that its use is not restricted
to any particular dimension, as instead are methods like the $\epsilon$-expansion, Montecarlo simulations
 or conformal field theory (CFT) techniques.
 Often one can obtain RG flow equations where
$d$ can be considered as an external tunable parameter and it is
thus possible to enquire how properties of the RG {}``theory space'' evolve
as the dimension is varied. In particular one can observe universality
classes, i.e. fixed-points of the RG flow, emerge or disappear at various
threshold or critical dimensions. The study of how $\mathbb{Z}_{2}$-universality
classes depend on $d$ is particularly interesting since it offers
the possibility of observing how the fixed-points structure of theory
space interpolates between the upper critical dimension $d=4$, where
there is only the Gaussian fixed-point, and $d=2$ where one expects,
from the CFT analysis, an infinite discrete sequence of fixed-points
describing multi-critical universality classes of arbitrary order.
Indeed this is what we are able to observe.
\\

\paragraph*{Flow equation for the effective potential.}

The effective average action $\Gamma_{k}[\varphi]$ is a generalization
of the standard effective action that depends on the infrared
scale $k$ \cite{Berges_Tetradis_Wetterich_2002}. The effective potential $V_{k}(\varphi)$ is found
by evaluating the effective average action at a constant field configuration
$\varphi(x)\equiv\varphi$ where $\Gamma_{k}[\varphi]=(\int d^d x) V_{k}(\varphi)$.
In this approximation theory space is projected into
the infinite dimensional functional space of effective potentials
and the RG flow is represented by an exact equation
that describes the scale dependence of the effective potential \cite{Wetterich_1993}.
In terms of dimensionless variables $\varphi=Z_{k}^{-1/2}k^{d/2-1}\tilde{\varphi}$ and $V(\varphi)=k^{d}\tilde{V}(\tilde{\varphi})$,
where $Z_{k}$ is the running wave-function renormalization constant,
the exact flow equation for the effective potential reads \cite{Ballhausen_Berges_Wetterich_2004}:
\begin{equation}
\partial_{t}\tilde{V}_{k}(\tilde{\varphi})+d\tilde{V}_{k}(\tilde{\varphi})-\frac{d-2+\eta_{k}}{2}\tilde{\varphi}\,\tilde{V}_{k}'(\tilde{\varphi})=c_{d}\frac{1-\frac{\eta_{k}}{d+2}}{1+\tilde{V}_{k}''(\tilde{\varphi})}\,,\label{2}
\end{equation}
with $c_{d}^{-1}=(4\pi)^{d/2}\Gamma(d/2+1)$.
Note that equation (\ref{2}) is a non-linear partial differential
equation (PDE).
The scale dependent anomalous dimension, introduced in (\ref{2}), is
defined in terms of the wave-function renormalization by $\eta_{k}=-\partial_{t}\log Z_{k}$;
it can be obtained from the dimensionless effective potential using
the following relation \cite{Ballhausen_Berges_Wetterich_2004}:
\begin{equation}
\eta_{k}=c_{d}\frac{[\tilde{V}_{k}'''(\tilde{\varphi}_{0})]^{2}}{[1+\tilde{V}_{k}''(\tilde{\varphi}_{0})]^{4}}\,.\label{2.1}
\end{equation}
In (\ref{2.1}) $\tilde{\varphi}_{0}$ is the minimum of the dimensionless
effective potential, i.e. $\tilde{V}_{k}'(\tilde{\varphi}_{0})=0$.

An interesting feature of equation (\ref{2}) is that it is valid for
any value of $d$; in principle it contains information about universality
classes in any dimension. In particular we expect to find non-trivial
behaviors in $d=3$, where equation (\ref{2}) has been extensively
studied \cite{Litim_2001}, and in $d=2$ where one expects a rich
behavior, since the CFT analysis implies the existence of an infinite
number of non-trivial RG fixed-points \cite{CFT}.
\\

\paragraph*{Scaling Solutions.}

Scaling solutions are solutions of $\partial_{t}\tilde{V}_{*}(\tilde{\varphi})=0$
and correspond to RG fixed-points in the functional space of effective
potentials. Every scaling solution, together with its domain of attraction,
defines a different universality class. 

We start our analysis by considering the case where the anomalous dimension
is set to zero $\eta=0$. From (\ref{2}) we see that a scaling
solution satisfies the following ordinary differential equation (ODE):
\begin{equation}
-d\tilde{V}_{*}(\tilde{\varphi})+\frac{d-2}{2}\tilde{\varphi}\,\tilde{V}_{*}'(\tilde{\varphi})+c_{d}\frac{1}{1+\tilde{V}_{*}''(\tilde{\varphi})}=0\,.\label{3}
\end{equation}
The $\mathbb{Z}_{2}$-symmetry of the effective potential requires
that its first derivative vanishes at the origin $\tilde{V}_{*}'(0)=0$;
(\ref{3}) then implies $\tilde{V}_{*}(0)=\frac{c_{d}/d}{1+\tilde{V}_{*}''(0)}$.
Since equation (\ref{3}) is a second order non-linear ODE, we need
to use numerical methods to solve it \footnote{All the numerical analysis has been performed employing standard routines for solving ODE present in symbolic manipulation software packages.}. It's easy to set up the initial
value problem as a function of the parameter $\sigma=\tilde{V}_{*}''(0)$ using the two
initial conditions just given.
\begin{figure*}
\begin{centering}
\includegraphics[scale=1]{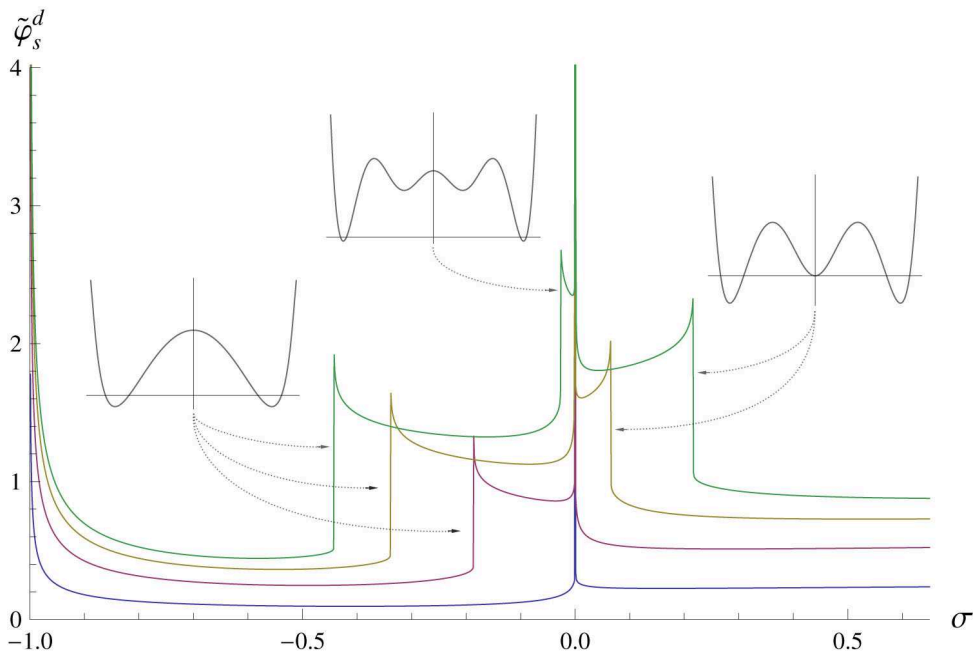}
\par\end{centering}
\caption{The function $\tilde{\varphi}_{s}^{d}(\sigma)$ for (starting from
below) $d=4,3,2.6,2.4$. In $d=4$ one finds only the Gaussian scaling solution,
represented by  a spike at the origin. As the dimension is lowered
new spikes emerge from the Gaussian spike, alternatively on the left
and on the right; these correspond to multi-critical scaling solutions
of increasing degree. In particular, in $d=3$ one observes a spike
at negative $\sigma$, corresponding to the Wilson-Fisher scaling
solution; in $d=2.6$ the spike at positive $\sigma$ corresponds to the tri-critical Ising scaling solution,
while in $d=2.4$ the newer spike corresponds to the tetra-critical scaling solution.
In the insets we show the general form of the dimensionless effective potentials
obtained by integrating (\ref{3}) using, as initial conditions, the
positions of the relative spikes, i.e. the values $\sigma_{*,i}$.}
\end{figure*}

One immediately observes that for most values of the parameter
$\sigma$ the solution ends up in a singularity at a finite value
of the dimensionless field. For every $d$ and $\sigma$ we can call
this value $\tilde{\varphi}_{s}^{d}(\sigma)$, in this way defining
a function \cite{Morris_1994a}. Requiring a scaling solution to be
well defined for any $\tilde{\varphi}\in\mathbb{R}$ restricts the
admissible initial values of $\sigma$ to a discrete set $\{\sigma_{*,i}^{d}\}$
(labeled by $i$). One can now construct a numerical plot of the function
$\tilde{\varphi}_{s}^{d}(\sigma)$ to find the $\sigma_{*,i}$ as
those values where the function $\tilde{\varphi}_{s}^{d}(\sigma)$
has a ``spike'', since a singularity in $\tilde{\varphi}_{s}^{d}(\sigma)$ implies that the relative
scaling solution, obtained by integrating the ODE (\ref{3}), is a
well defined function for every $\tilde{\varphi}\in\mathbb{R}$. For
any $d$, the function $\tilde{\varphi}_{s}^{d}(\sigma)$ gives
us a snapshot of theory space, where dimensionless  effective
potentials are parametrized by $\sigma$ and where RG fixed-points,
i.e. scaling solutions, appear  as spikes. By studying
$\tilde{\varphi}_{s}^{d}(\sigma)$ we will be able to follow the evolution
of universality classes as we vary the dimension.

We can start by studying the function $\tilde{\varphi}_{s}^{d}(\sigma)$
for $d=4$. One finds only one spike at $\sigma_{*,1}=0$ corresponding
to the Gaussian scaling solution $\tilde{V}_{*}(\tilde{\varphi})=\frac{c_{d}}{d}$
(the singularity at $\sigma=-1$ is due to the structure of equation
(\ref{3}) and does not correspond to any scaling solution). The function
$\tilde{\varphi}_{s}^{4}(\sigma)$ is shown in Figure 1. We find the
same qualitative result for any $d\geq4$ as expected by the fact that four is
the upper critical dimension for the Ising (or Wilson-Fisher) universality class.

As we decrease the dimension from $d=4$ we observe a new spike
branching to the left of the Gaussian spike: this corresponds to
the Ising scaling solution. As we continue to lower
$d$ the spike moves to the left and for $d=3$ the function $\tilde{\varphi}_{s}^{3}(\sigma)$
looks as in Figure 1. As one expects, the value of $\sigma_{*,2}$ at which we observe
the Ising spike is negative, indicating that the relative scaling
solution obtained by integrating the fixed-point equation (\ref{3})
is concave at the origin.

For values of the dimension lower than three the function $\tilde{\varphi}_{s}^{d}(\sigma)$
becomes very interesting. Starting at $d=3$ a new spike branches
from the Gaussian one, this time to the right: this corresponds
to the tri-critical Ising universality class, for which this is the
upper critical dimension. When we reach $d=2.6$ this spike is clearly
visible, as is shown in Figure 1. In $d=2.4$ another spike has already
emerged, this time to the left of the Gaussian spike: this corresponds
to the tetra-critical Ising universality class.
In the insets of Figure 1 we show the general form of the scaling solutions obtained
by integrating equation (\ref{3}) using, as initial conditions, the
positions of the spikes, i.e. the values $\sigma_{*,i}$ for $i=2,3,4$.
One sees that these solutions indeed correspond to dimensionless effective
potentials with two, three and four minima.

\begin{figure}
\begin{centering}
\includegraphics[scale=0.68]{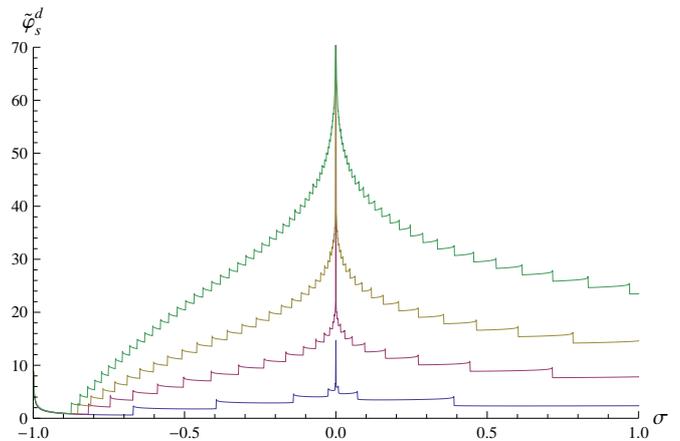}
\par\end{centering}
\caption{The function $\tilde{\varphi}_{s}^{d}(\sigma)$ for values of the
dimension approaching two; (form bottom) $d=2.1,2.01,2.003,2.001$.
One observes the increasing number of spikes corresponding to multi-critical
scaling solution of increasing degree. An arbitrary number of spikes
can be observed as $d\rightarrow2$, the only constraint being the
numerical resolution at which the function $\tilde{\varphi}_{s}^{d}(\sigma)$
is investigated.}
\end{figure}

As we lower further the dimension new spikes emerge alternatively
on the left or on the right of the Gaussian spike. These correspond
to multi-critical potentials of increasing order. In particular, one observes that
the $i$-th multi-critical scaling solution appears at the critical
dimension $d_{c,i}$;
these can be understood as those dimensions where
new operators become relevant as $d$ is lowered. The dimension of
the operator $\varphi^{2i}$ is $2i(\frac{d}{2}-1)$, so the relative
coupling $\lambda_{2i}$ scales as $d-2i(\frac{d}{2}-1)$. Setting
this to zero gives $d_{c,i}=\frac{2i}{i-1}$; for integer values of $i=1,2,3,4,...$
we find the following sequence of critical dimensions $d_{c,i}=\infty,4,3,\frac{8}{3},\frac{5}{2},\frac{12}{5},\frac{7}{3},\frac{16}{7},\frac{9}{4},...$
Infinity is the upper critical dimension for the Gaussian universality
class, four is the upper critical dimension for the Ising universality
class, while three is the upper critical dimension the tri-critical Ising
universality class and so on. Note that the critical dimensions $d_{c,i}$
accumulate at $d=2$: this is a first hint that in two dimensions
there are infinitely many different universality classes. By plotting
the function $\tilde{\varphi}_{s}^{d}(\sigma)$ with the appropriate
resolution one can check that the new spikes emerge precisely at the dimensions
$d_{c,i}$. It will be intresting to analytically confirm these results,
as was done for the Polchinski analogue of (\ref{3}) in \cite{Felder_1987}.

As shown in Figure 2, when we continue to decrease $d$ towards two, the
number of spikes of $\tilde{\varphi}_{s}^{d}(\sigma)$ grows as expected. 
In $d=2$ we  "lose track" of the multi-critical scaling solutions; this is related to the fact
that the first derivative term on the lhs of (\ref{3}) vanishes, making the equation exactly
integrable.\\

\paragraph*{Sine-Gordon model.}

To integrate the fixed-point equation (\ref{3}) when $d=2$, we recast it in the following form:
\begin{equation}
\tilde{V}_{*}''(\tilde{\varphi})=-\frac{d}{d\tilde{V}_{*}}\left[\tilde{V}_{*}-\frac{1}{8\pi}\log\tilde{V}_{*}\right]\,.\label{4}
\end{equation}
Equation (\ref{4}) can be interpreted as Newton's equation where
$\tilde{V}(\tilde{\varphi})\leftrightarrow x(t)$ \cite{Morris_1994b};
its solution is thus implicitly given by
\begin{equation}
\tilde{\varphi}=\int\frac{d\tilde{V}_{*}}{\sqrt{2\left(\frac{c_{d}/d}{1+\sigma}-\tilde{V}_{*}\right)+\frac{1}{4\pi}\log\frac{(1+\sigma)\tilde{V}_{*}}{c_{d}/d}}}\,.\label{5}
\end{equation}
Since the {}``potential'' $\tilde{V}_{*}-\frac{1}{8\pi}\log\tilde{V}_{*}$
in (\ref{4}) is convex and bounded from below, the solutions (\ref{5})
are periodic functions of $\tilde{\varphi}$; the period depends on
$\sigma$ and the solution with zero period is the Gaussian scaling solution. We thus see
that in $d=2$ equation (\ref{3}) has a continuum of periodic solutions.
These are related to the critical sine-Gordon model \cite{Nandori_Nagy_Sailer_Trombettoni_2009}.
One can expand the dimensionless effective potential in Fourier series
$\tilde{V}_{*}(\tilde{\varphi})=\sum_{n=1}^{\infty}v_{n,*}\cos(n\beta_{*}\tilde{\varphi})$
and check that indeed one finds the fixed-point value
$\beta_{*}=\sqrt{8\pi}$ corresponding to the Coleman point.
In the sine-Gordon model the wave-function renormalization is an essential coupling
related to the vortex-fugacity; one can then use (\ref{2}) and (\ref{2.1}) to obtain beta functions
exibiting the Kosterlitz-Thouless-Berezinski phase transition \cite{Nagy_Nandori_Polonyi_Sailer_2009}.
\\
\begin{figure}
\begin{centering}
\includegraphics[scale=0.68]{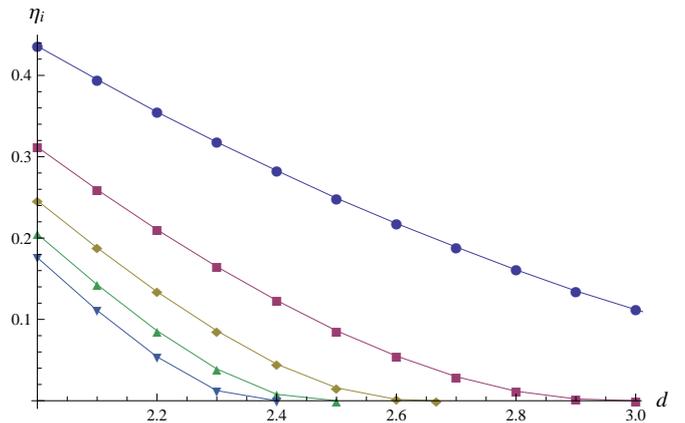}
\par\end{centering}
\caption{The anomalous dimensions $\eta_{i}$ of the
first five multi-critical scaling solutions as a function of $d$.}
\end{figure}

\paragraph*{Anomalous dimension.}

In order to follow the evolution of scaling solutions down to two dimensions,
we turn to consider the case of non-vanishing anomalous dimension
$\eta\neq0$. For every scaling solution we will determine the allowed values of 
$\eta$ by solving, self-consistently, the fixed-point equation
\begin{equation}
-d\tilde{V}_{*}(\tilde{\varphi})+\frac{d-2+\eta}{2}\tilde{\varphi}\,\tilde{V}_{*}'(\tilde{\varphi})+c_{d}\frac{1-\frac{\eta}{d+2}}{1+\tilde{V}_{*}''(\tilde{\varphi})}=0\,,\label{5.1}
\end{equation}
together with (\ref{2.1}). Since now the second
term on the lhs of (\ref{5.1}) is non-zero even when $d=2$, due
to a non-vanishing anomalous dimension, we expect to find
a discrete set of scaling solutions for every $d\geq2$.
\begin{figure}
\begin{centering}
\includegraphics[scale=0.68]{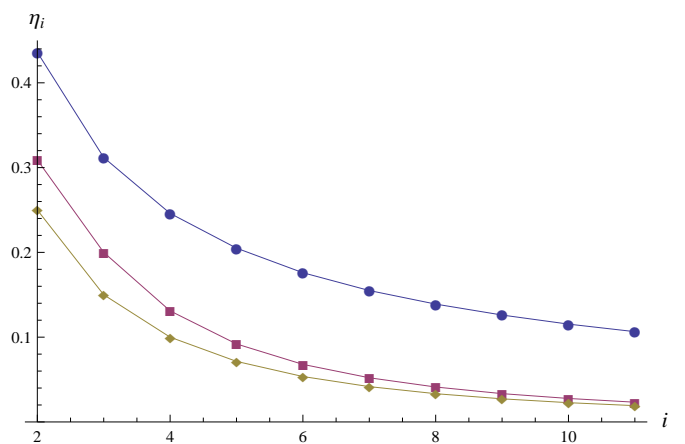}
\par\end{centering}
\caption{The anomalous dimensions $\eta_{i}$ of the first ten multi-critical
scaling solutions in $d=2$. We show the estimates of our work (upper curve)
together with results of the derivative expansion to order $\partial^{2}$
(middle curve) and the exact results from the CFT analysis (lower
curve).}
\end{figure}
Following the $\eta=0$ case, we define the functions $\tilde{\varphi}_{s}^{d,\eta}(\sigma)$
and identify the discrete values $\{\sigma_{*,i}^{d,\eta}\}$ for
which the solutions of (\ref{5.1}) are well defined for every $\tilde{\varphi}\in\mathbb{R}$.
The functions $\tilde{\varphi}_{s}^{d,\eta}(\sigma)$, for $\eta\neq0$, turn out to be qualitatively
similar to their counter-parts $\tilde{\varphi}_{s}^{d}(\sigma)$ studied previously.

We proceed as follows: we fix $d$ and we start with an ansatz for
$\eta$; we compute $\tilde{\varphi}_{s}^{d,\eta}(\sigma)$ from which
we find the values $\{\sigma_{*,i}^{d,\eta}\}$; we use them to solve
numerically the ODE (\ref{5.1}) to obtain the relative scaling solutions;
we estimate the anomalous dimension by employing (\ref{2.1}); we use
this value as the ansatz for the next iteration until we converge
to a self-consistent solution of (\ref{5.1}).
Using this procedure we were able to find scaling solutions
of (\ref{5.1}) together with the relative anomalous dimensions. In
particular we started at $d=4$ where $\eta=0$ and we followed the
scaling solutions down to $d=2$ by making steps in $d$ of size $0.1$.
At every critical $d_{c,i}$ we were able to follow the emerging multi-critical
scaling solution. This allowed us to calculate the anomalous dimensions
$\eta_{i}$ of the multi-critical fixed-points as a function of $d$.
In Figure 3 we show the anomalous dimensions of the first five multi-critical
scaling solutions in the range $2\leq d\leq3$.
In particular, in $d=2$ our fixed-points can be directly related
to the multi-critical unitary minimal models that are found in CFT.

In $d=2$ we compared our estimates for the anomalous dimensions
with those found by Morris \cite{Morris_1994b}, using the derivative
expansion to order $\partial^{2}$, and with those obtained from the
exact relation $\eta_{i}=\frac{3}{(i+1)^{2}+i+1}$, obtained by CFT
techniques \cite{CFT}. We make the comparison for the first
ten multi-critical scaling solutions in Figure 4. Due to our elementary
estimate (\ref{2.1}), our values systematically
over-estimate both the $\partial^{2}$ and exact values; still it
is astonishing that we correctly reproduce the overall qualitative
picture. We stress that we where able to observe an arbitrary number
of scaling solutions and not only the first ten as in \cite{Morris_1994b}.
Our RG analysis predicts an infinite sequence of multi-critical scaling
solutions of increasing degree, with anomalous dimensions decreasing
monotonically with $i$. It is important to remark that nowhere did we
assume conformal invariance, but just  translation,
rotation, $\mathbb{Z}_{2}$ and RG scale invariance! All our conclusions where drawn just from
the analysis of the simple ODE (\ref{5.1}). 

Our analysis can be seen as a RG validation of the correspondence between Landau-Ginsburg actions and minimal models \cite{CFT2};
correspondence that cannot be seen at the level of beta functions alone, but requires at least a functional truncation of the RG theory space, since only in this case it is possible to distinguish spurious fixed-points from real ones and thus to disentangle the rich fixed-point structure observed near $d=2$. This is probably one reason why the $\epsilon$-expansion  fails to correctly detect minimal models in two dimensions \cite{Howe:1989hc}.  

In view of the results of this paragraph,
and of those of \cite{Morris_1994b}, one can expect to reproduce
CFT results at the quantitative level by employing functional RG techniques.
Obviously the payback is that the functional RG analysis can be done independently
and continuously in the dimension.
\\

\paragraph*{Critical exponents.}

Once we have found a scaling solution, together with its
anomalous dimension, we can obtain the other critical exponents by
studying linear perturbations around it. An easy way to do this \cite{Litim_2001}
is by expanding the potential in a Taylor series $\tilde{V}(\tilde{\varphi})=\sum_{n=1}^{N_{tr}}\frac{\lambda_{2n}}{(2n)!}\tilde{\varphi}^{2n}$
and to transform the PDE (\ref{4}) in a system of $N_{tr}$ coupled
ODE for the beta functions $\tilde{\beta}_{2n}=\partial_{t}\tilde{\lambda}_{2n}$
of the dimensionless couplings $\tilde{\lambda}_{2n}=k^{d-2n\left(\frac{d}{2}-1+\frac{\eta}{2}\right)}\lambda_{2n}$
(note that we consider here $\eta$ as an input from the preceding analysis). RG fixed-points now correspond to solutions of the
following system of algebraic equations:
\begin{equation}
\tilde{\beta}_{2n}=0\qquad\qquad n=1,...,N_{tr}.\label{6}
\end{equation}
As we vary the number $N_{tr}$ of couplings in our truncation we
usually find more and more solutions to the system (\ref{6}). Most
of these are spurious fixed-points: only with the previous analysis
of the scaling solutions it is possible to understand which of these
do correspond to admissible solutions of the fixed-point equation
(\ref{5.1}) and which instead correspond to singular solutions.
Since $\tilde{\lambda}_{2}=\sigma$, the values $\{\sigma_{*,i}^{d,\eta}\}$ that
we previously obtained for given $d$ and $\eta$ now serve as a
{}``map'' to correctly spot the desired solutions of the algebraic
system (\ref{6}). We clearly understand that the spurious solutions
of (\ref{6}) correspond to solutions of (\ref{5.1}) which end
up in a singularity \cite{Morris_1994a}.
\begin{figure*}
\begin{centering}
\includegraphics[scale=1]{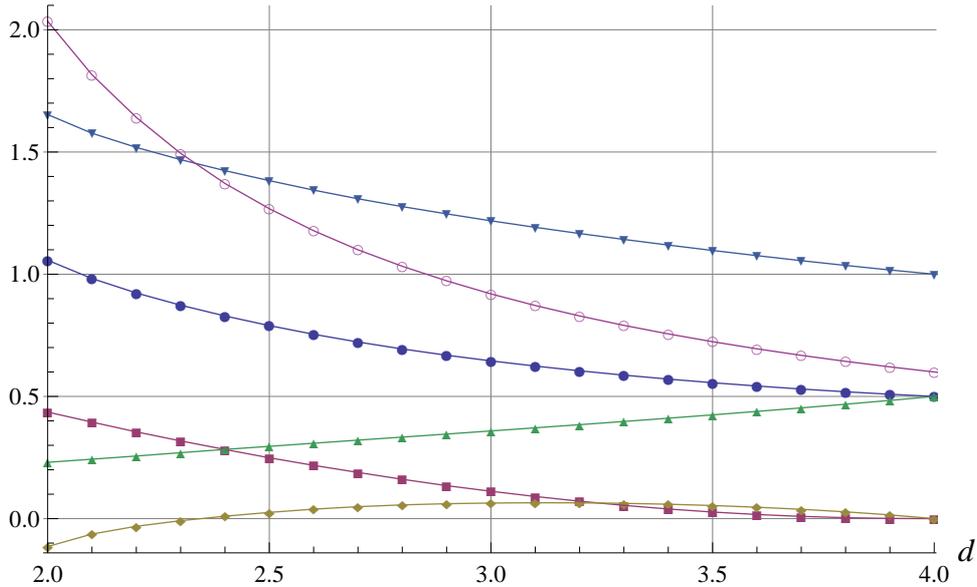}
\par\end{centering}
\caption{Critical exponents for the Ising universality class in the range $2\le d\le4$.
From top-left $\delta_{2}(d)/5$ (empty circles), $\gamma_{2}(d)$ (downward
triangles), $\nu_{2}(d)$ (filled circles), $\eta_{2}(d)$ (squares), $\beta_{2}(d)$
(upward triangles) and $\alpha_{2}(d)$ (rhombuses).
}
\end{figure*}

\begin{table}
\begin{centering}
\begin{tabular}{|c|c|c|c|c|}
\cline{2-5} 
\multicolumn{1}{c|}{} & \multicolumn{2}{c|}{$d=2$} & \multicolumn{2}{c|}{$d=3$}\tabularnewline
\cline{2-5} 
\multicolumn{1}{c|}{} & {\small this work} & {\small exact} & {\small this work} & {\small{} world best}\tabularnewline
\hline 
$\eta$ & $0.436$ & $0.25$ & $0.11$ & $0.036$\tabularnewline
\hline 
$\nu$ & $1.05$ & $1$ & $0.65$ & $0.63$\tabularnewline
\hline 
$\alpha$ &-0.11& $0$ & $0.06$ & $0.11$\tabularnewline
\hline 
$\beta$ & $0.23$ & $0.125$ & $0.36$ & $0.33$\tabularnewline
\hline 
$\gamma$ & $1.65$ & $1.75$ & $1.22$ & $1.24$\tabularnewline
\hline 
$\delta$ & $10.17$ & $15$ & $4.60$ & $4.79$\tabularnewline
\hline
\end{tabular}
\par\end{centering}
\caption{Calculated critical exponents for the Ising universality class in
two and three dimensions compared to exact results \cite{CFT}
and {}``world best'' estimates \cite{Pelissetto_Vicari_2002}.}
\end{table}

It is useful to proceed as follows. Since the beta functions $\tilde{\beta}_{2n}$
are linear in the couplings $\tilde{\lambda}_{2n}$, we can iteratively solve
(\ref{6}) to find the couplings $\tilde{\lambda}_{2n}(\sigma)$ in terms of $\tilde{\lambda}_{2}=\sigma$.
Inserting these in $\tilde{\beta}_{2n+2}=0$ makes it a polynomial equation
in $\tilde{\lambda}_{2}=\sigma$. This can be solved numerically;
from the solutions we find, we pick the one that best
approximates the value $\sigma_{*,i}$ relative to the scaling solution
we are Taylor expanding. Then we obtain the fixed-point values for all
the dimensionless couplings $\tilde{\lambda}_{2n}(\sigma_{*,i})$.
To determine the critical exponents,
we just need to calculate the stability matrix and evaluate it at $\tilde{\lambda}_{2n}(\sigma_{*,i})$:
$M_{nm}=\left.\frac{\partial\tilde{\beta}_{2n}}{\partial\tilde{\lambda}_{2m}}\right|_{*}$.

We did this analysis for the Ising universality class. In this
case the eigenvalues of $M_{nm}$ are such that $\Lambda_{1}(d)<0<\Lambda_{2}(d)<\Lambda_{3}(d)<\cdots$
for any $2\leq d\leq4$. By varying the number of couplings $N_{tr}$
in our truncation we were able to obtain convergent estimates
for the eigenvalues 
\footnote{We repeated the analysis by performing a Taylor expansion around the minimum of the dimensionless effective
potential.}.
We obtained estimates for the correlation length critical exponent
$\nu_{2}(d)=-1/\Lambda_{1}(d)$ as a function of the dimension. We
show the results of our analysis in Figure 5, together with $\eta_{2}(d)$
and the other critical exponents that we obtained by employing standard scaling
relations \cite{Pelissetto_Vicari_2002}.
In two and three dimensions
we found the results listed in Table 1.
Our estimates for the exponents
$\nu_{2}(d)$ and $\gamma_{2}(d)$ turned out particularly good, both in $d=2$
and $d=3$. $\delta_{2}(d)$ and $\beta_{2}(d)$ are
well estimated in $d=3$ but not so well in $d=2$, while the value of $\alpha_{2}(d)$ is poor in both cases. 
The quantitative estimate for $\eta_{2}(d)$, as we observed earlier, is more crude due
to the elementary ansatz (\ref{2.1}).
\\

\paragraph*{Discussion and outlook.}

In this paper we studied how universality classes of scalar theories with $\mathbb{Z}_{2}$-symmetry
evolve with the dimension $d$. Using functional RG techniques based on the effective average action,
we were able to follow, continuously with $d$, the evolution of RG fixed-points (represented by scaling solutions)
through the functional theory space of effective potentials.
Even if all our analysis was based on the study of
a simple ODE, we were able to observe a very rich behavior.
Above four dimensions we found only the Gaussian universality
class; at $d=3$ the Ising or Wilson-Fisher universality class appeared;
while in two dimensions we found a countable infinity of multi-critical
fixed-points corresponding to the unitary minimal models of CFT. 
More importantly, we presented the continuum evolution with $d$ of these universality
classes; we observed a succession of critical dimensions $d_{c,i}$
(that accumulate at $d=2$) where new universality classes branched from
the Gaussian one.
The picture that we presented captured all possible
critical behavior associated to $\mathbb{Z}_{2}$-symmetry in any
dimension $d\geq2$.
We were also able to give quantitative estimates for critical exponents
as a function of $d$.

It will be very interesting to find real physical systems,
of effective fractal dimension $d$, where to observe the phase transitions
related to these universality classes, and to compare our critical exponents with experiments.
Otherwise these systems could be studied on the lattice \cite{Vezzani_2005}. 
Extensions of the approach presented in this paper can be very
fruitful to deal with the general problem of classifying
universality classes, and to study how these depend continuously on $d$.
\\

\paragraph*{Acknowledgments}

We would like to thank R. Percacci, O. Zanusso, G. D'Odorico and A. Trombettoni for useful and stimulating discussions.


\begin{thebibliography}{10}

\bibitem{Berges_Tetradis_Wetterich_2002}J. Berges, N. Tetradis and C. Wetterich,
Phys.\ Rept.\  363 (2002) 223, hep-ph/0005122.

\bibitem{Wetterich_1993}C. Wetterich, Phys. Lett. B 301 (1993) 90.

\bibitem{Ballhausen_Berges_Wetterich_2004}H. Ballhausen, J. Berges
and C. Wetterich, Phys. Lett. B 582 (2004) 144, hep-th/0310213.


\bibitem{Litim_2001}D.F. Litim, Phys. Rev. D 64 (2001) 105007, hep-th/0103195.


\bibitem{CFT}
A.A. Belavin, A.M. Polyakov and A.B. Zamolodchikov, Nucl.\ Phys.\ B {\bf 241} (1984) 333.

\bibitem{Morris_1994a}T.R. Morris, Phys. Lett. B 334 (1994) 355,
hep-th/9405190.

\bibitem{Felder_1987}G. Felder, Comm. Math. Phys. 111 (1987) 101.



\bibitem{Nandori_Nagy_Sailer_Trombettoni_2009}I. Nandori, S. Nagy, K. Sailer and A. Trombettoni, 
Phys.\ Rev.\ D 80 (2009) 025008, arXiv:0903.5524 [hep-th].

\bibitem{Nagy_Nandori_Polonyi_Sailer_2009}S. Nagy, I. Nandori, J. Polonyi and K. Sailer,
Phys.\ Rev.\ Lett.\  102 (2009) 241603, arXiv:0904.3689 [hep-th].

\bibitem{Morris_1994b}T.R.Morris, Phys. Lett. B 345 (1995) 139, hep-th/9410141.

\bibitem{CFT2}
A.B. Zamolodchikov, Sov.\ J.\ Nucl.\ Phys.\  {\bf 44} (1986) 529 [Yad.\ Fiz.\  {\bf 44} (1986) 821].

\bibitem{Howe:1989hc}
P.S. Howe and P.C. West, Phys.\ Lett.\ B {\bf 223} (1989) 371.


\bibitem{Pelissetto_Vicari_2002}A. Pelissetto and E. Vicari, Phys.
Rept. 368 (2002) 549, cond-mat/0012164.

\bibitem{Vezzani_2005}A. Vezzani, J Phys. A 36 (2003) 1593-1604, cond-mat/0212497.

\end{thebibliography}
\end{document}